\documentclass[twocolumn,showpacs,preprintnumbers,amsmath,amssymb]{revtex4}
\usepackage{graphicx}% Include figure files
\usepackage{dcolumn}% Align table columns on decimal point
\usepackage{bm}% bold math

\newcommand {\be}{\begin{equation}}
\newcommand {\ee}{\end{equation}}
\newcommand {\bea}{\begin{eqnarray}}
\newcommand {\eea}{\end{eqnarray}}
\newcommand {\bem}{\begin{displaymath}}
\newcommand {\eem}{\end{displaymath}}
\newcommand {\f}{\frac}
\begin{document}

\preprint{ }

\title{Current sharing between superconducting film and normal metal}% Force line breaks with \\
\author{ George A. Levin, Paul N. Barnes, and John S. Bulmer}
%\altaffiliation[Also at ]{Physics Department, XYZ University.}%Lines break %automatically or can be forced with \\
%\author{ }
% \email{Second.Author@institution.edu}
\affiliation{Propulsion Directorate, Air Force Research Laboratory,  1950 Fifth Street, Bldg. 450, Wright-Patterson Air Force Base, OH 45433}
%
%\author{ }
%\homepage{http://www.Second.institution.edu/~Charlie.Author}
%\affiliation{ Department of Materials Science and Engineering, The Ohio State %University, Columbus, OH 43210}
%Second institution and/or address\\
%This line break forced% with \\
%}%

\date{\today}% It is always \today, today,
             %  but any date may be explicitly specified

\begin{abstract}
A two-dimensional model is introduced that describes current sharing between the superconducting and normal metal layers in configuration typical of YBCO-coated conductors. The model is used to compare the effectiveness of surround stabilizer and more conventional one-sided stabilizer. When the resistance of the interface between the superconductor and stabilizer is low enough, the surround stabilizer is less effective than the one-sided stabilizer in stabilizing a hairline crack in the superconducting film. \end{abstract}
%\keywords{Suggested keywords}%Use showkeys class option if keyword
                              %display desired
\maketitle

\section{\label{sec:level1}Introduction\protect}
Coated conductors are one of the most promising candidates for broader commercialization of high-$T_c$ superconductors. The architecture of coated conductors can be described as a "sandwich"  in which the $YBa_2Cu_3O_{7-x}$  (YBCO) superconducting film of about $1 \mu m$ thick is enclosed between a metal substrate of relatively high resistivity ($Ni-5\%W$ alloy, Hastelloy, or stainless steel) and a copper stabilizer layer\cite{Larbalestier,Schoop,Xie,Iijima, Usoskin1}.
The purpose of the stabilizer is to carry the transport current over the parts of the superconducting film that become temporarily or even permanently damaged or heated. An understanding of the processes and control of the characteristics of the current exchange between the superconductor and copper stabilizer is essential for developing effective stabilization of the conductor. 

In coated conductors the superconducting film is deposited on buffered metal substrate and the stabilizer is either soldered or electroplated on top of the YBCO film. Depending on the method by which the stabilizer is attached to the superconducting film, a thin layer of various chemical composition is formed as a resistive interface between them. The interface resistance is thought to be determined by a few "dead" (underdoped) unit cells of YBCO which has high normal state resistivity in the $c-$direction\cite{Angurel}. The spatial scale of the current exchange depends on the relationship between the interfacial resistance and the thickness and resistivity of the stabilizer. A one-dimensional model that describes such current exchange is well known. Here we have generalized it to a two-dimensional version. The model is then used to determine the effectiveness of a "surround stabilizer". 

Section 2 of this paper introduces a planar model of voltage distribution in the stabilizer, which is a two-dimensional generalization of the one-dimensional model that has been in use for some time\cite{Fang,Polak2,Cesnak,Usoskin,Fu,Stenvall,Takacs}.  In Section 3 we compare the effectiveness of two different types of stabilizers - "one-sided stabilizer"\cite{Larbalestier} and "surround stabilizer"\cite{Schoop,Xie}. Our conclusion is that for stabilization of hairline fractures in the superconducting layer the one-sided stabilizer is more effective than the surround stabilizer. This is true especially when the interface resistance is relatively low, so that the spatial scale of the current exchange is much smaller than the width of the conductor.

\section{\label{sec:level1}Planar model\protect}
Let us consider a layer of a normal metal in contact with a superconducting film, Fig. 1. 
Let $\bar {R}$ (with dimensionality $\Omega\; cm^2$) be the  resistivity of the interface between them, $d$ -
the thickness of the metal  and $\rho$- the resistivity of the metal. Here we consider only DC electric field and currents.
The density of current inside the normal metal
\be
\vec{j}=-\rho^{-1}\nabla\widetilde{V},
\ee
where $\widetilde{V}(x,y,z)$ is the electric potential. By virtue of current conservation, $div\vec{j}=0$, the potential is 
determined by the three-dimensional (3D) Laplace's equation:
\be 
\Delta\widetilde{V}=0.
\ee
At the superconductor –- metal interface, the potential is subject to the boundary condition that follows from the ohmic relationship between the current and voltage across the interface: 
\be
\left.  j_z\right |_{z=0}=-\rho^{-1}
\left.\frac{\partial{\widetilde{V}}}{\partial{z}}\right |_{z=0}=
-\frac{{V}-V_s}{\bar{R}}.
\ee
Here $V(x,y)\equiv \widetilde{V}|_{z=0}$ is the value of the potential just above the interface and $V_s$ is the potential of the superconducting film below the interface.  
The second boundary condition at $z=d$ depends on whether there are current leads attached to the stabilizer. If  there are no current leads  in the area of consideration, the boundary condition at $z=d$ is
\be
\left.\frac{\partial{\widetilde{V}}}{\partial{z}}\right |_{z=d}=0.
\ee
The effect of the current leads attached to the stabilizer can be accounted for by changing the boundary condition (4) as follows:
\be
-\rho^{-1}\left.\frac{\partial{\widetilde{V}}}{\partial{z}}\right |_{z=d}= j(x,y).
\ee
Here $j(x,y)$ is a given density of current injected through the leads. 

\begin{figure}
\includegraphics{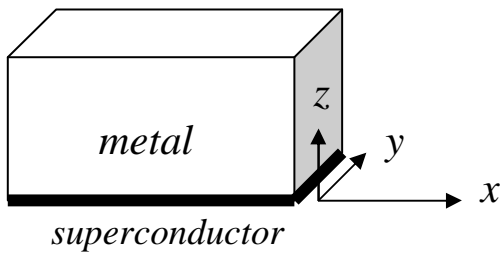}
\caption{\label{fig:} A sketch of a metal layer attached to a superconducting film.}
\end{figure}
\subsection{\label{sec:level2} Stabilizer}

A significant simplification of the model can be achieved in a practically important limit when the variation of the potential across the thickness of the stabilizer is small.
Then, we can use Taylor's expansion to approximate ${\widetilde{V}(x,y,z)}$ as follows:
\be
{\widetilde{V}(x,y,z)}\approx V(x,y) +\alpha z +\beta z^2.
\ee
The coefficients $\alpha$ and $\beta $ are determined by the boundary conditions (3)-(5). With the boundary conditions (3) and (4) (no current leads attached) the potential takes form:
\be
{\widetilde{V}(x,y,z)}=V +(V-V_s)\frac{z}{a} -  (V-V_s)\frac{z^2}{2da},
\ee
where $a$ is a characteristic length scale:
\be
a=\frac{\bar{R}}{\rho}.
\ee
The approximation (7) is valid as long as $d\ll a$. Another, less formal, way to interpret the criterion of validity of the planar approximation (7) is that the main resistance to current exchange is presented by the interface resistance, rather than by the vertical resistance of the stabilizer: $\rho d \ll \bar R $. Substituting the potential (7) in Eq. (2), and taking after the differentiation the limit $z\rightarrow 0$ we obtain for $V(x, y) $ the 2D Helmholtz equation:
\be
\Delta V -\kappa^2 (V-V_s)=0.
\ee
Here $\kappa \equiv \lambda^{-1} $ defines a screening length 
\be
\lambda = (ad)^{1/2}=\left (\frac{ d\bar R }{\rho }\right )^{1/2}.
\ee
The current can flow in the stabilizer only over distances of the order of $\lambda$. Everywhere else the stabilizer and superconducting film are equipotential and the current flows only in the superconducting layer.   

The average value of the potential 
\be 
V_{av}(x,y)=\frac{1}{d}\int_0^d \widetilde{V}(x,y,z)dz
\ee
is given by
\be
V_{av}(x,y)-V_s=(V-V_s)(1+\frac{d}{3a}),
\ee
and as long as $d\ll a$, the variation of the potential across the thickness of the metal can be neglected when we estimate losses and other average quantities.

With current leads attached to the stabilizer we use the boundary condition (5), instead of the condition (4), and the potential takes form:
\be
{\widetilde{V}(x,y,z)}=V +(V-V_s)\frac{z}{a} -  (V-V_s)\frac{z^2}{2da}-\rho j \frac{z^2}{2d}. 
\ee
Then, the Eq.(2) morphs into 2D inhomogeneous Helmholtz equation for $V(x,y)$:
\be
\Delta V -\kappa^2 (V-V_s)=\frac{\rho}{d}j(x,y).
\ee
The applicability of the approximation (13) requires  
\be
\rho j\sim (V-V_s)/a.
\ee
Taking into account Eqs. (3) and (8) it means that the density of current injected into the stabilizer from the leads should not substantially exceed the density of current exchanged between the stabilizer and the superconductor:
\be
j\sim \frac{V-V_s}{\bar{R}}.
\ee
This in turn means that the size of the footprint of the current lead should not be much smaller than the screening length $\lambda$.

\subsection{\label{sec:level2} Superconductor}

The equations describing the potential in the stabilizer have to be supplemented by the equations for the potential $V_s$ in the superconducting film. It is reasonable to describe the superconducting film with thickness much smaller than that of the stabilizer by two-dimensional density of current $\vec{J_s}\;(A/cm)$ subject to current conservation condition:
\be
\nabla\cdot\vec{J_s}=-\left.  j_z\right |_{z=0}=\frac{{V}-V_s}{\bar{R}}.
\ee
Here we have used Eq. (3). A constituent relation for the superconductor can be written as $\vec{E}\equiv r(J_s)\vec{J_s}$, which translates into a non-linear equation for the potential:
\be
\nabla [r^{-1}(|\nabla V_s|)\nabla V_s]=-\frac{{V}-V_s}{\bar{R}}.
\ee
Using a power law dependence\cite{Friesen}
\be
r(J_s)=\f{E_0}{J_c}\f{|J_s|^{n-1}}{|J_c|^{n-1}},
\ee
Eq. (18) takes the form:
\be
\nabla \left [\nabla V_s \left (\f{|\nabla V_s|}{E_0}\right )^{(1-n)/n}\right ]=\Lambda^{-2}(V-V_s).
\ee
Here $J_c$ is the critical current density and $E_0=1 \mu V/cm$. The screening length in the superconductor 
\be
\Lambda =\left (\f{\bar{R}J_c}{E_0}\right )^{1/2},
\ee
along with that in the stabilizer - Eq.(10), determines the distances over which the current exchange between superconductor and stabilizer takes place.

Measurements carried out on currently manufactured coated conductors\cite {Polak} indicate that their interface resistivity is $\bar{R}\sim 50\; n\Omega\;cm^2$. At $77\; K$ the  resistivity of copper is $\rho\approx 0.2\times 10^{-6}\Omega\; cm$. The thickness of the stabilizer is $d\sim 20 - 40 \mu m$. Thus, the characteristic length $a$ defined by Eq. (8) is
\be
a\sim 2.5 \; mm,
\ee
and the condition of applicability of the planar model $d\ll a$ is well satisfied. The screening length in the stabilizer is
\be
\lambda =(da)^{1/2}\sim 300\;\mu m.
\ee
With the critical current density in coated conductors $J_c\sim 200\; -\; 400 A/cm$, the corresponding screening length in the superconductor is
\be
\Lambda\sim 3\;-\;4 \;cm.
\ee
A huge difference in these screening lengths can be understood as the difference between the sheet resistance of copper $\rho/d\sim 10^{-4}\;\Omega$ and the equivalent quantity $E_0/J_c\sim 10^{-8}\;\Omega$ in the superconducting film.  Thus, the effects of the current exchange in the stabilizer manifest itself over much shorter distances than those in the superconductor. The potential difference between superconductor and stabilizer tends to be eliminated over distances of the order of $\lambda$ because the stabilizer adjusts its potential to that of the superconductor. 

The length scales of the potential distribution in the superconducting films tend to be large, of the order of $bn$ or $bn^{1/2}$, in the transverse and longitudinal direction respectively,  where $b$ is the size of a defect or an obstacle to current and $n\sim 20-40$ is the exponent in Eq. (19)\cite{Friesen}. Therefore, the details of the current exchange that takes place on the spatial scale of the order of $\lambda$ in many situations can be accurately described by assuming that the potential of the superconductor is constant or piecewise constant. Generally, as a good approximation, one can neglect the right-hand side in Eq. (20) and determine the potential $V_s$ as if there were no current exchange with the stabilizer\cite{Friesen}. Then, the solution of Eq.(9) is given by
\be
V(x,y)=-\kappa^{2}\int G(x-x^{\prime},y-y^{\prime})V_s(x^{\prime},y^{\prime})dx^{\prime} dy^{\prime},
\ee
where $ G(x-x^{\prime},y-y^{\prime})$ is the Green's function of the Helmholtz equation. The accuracy of this approximation is of the order of $\lambda/\Lambda$ and can be improved by iteration.

\subsection{\label{sec:level2} Energy losses}
There are several general statements about energy losses that follow from the planar model. The power dissipation has three components. First is the dissipation in the stabilizer:
\be
Q_{st}=\int \rho^{-1}|\nabla\widetilde{V}|^{2}dxdydz\approx 
\f{d}{\rho }\int |\nabla V|^{2}dxdy.
\ee
Second is the power dissipation in the interface itself - Eq. (3):
\be
Q_{int}=-\int \left.  j_z\right |_{z=0}(V-V_s)dxdy=\int
\f{(V-V_s)^2}{\bar{R}}dxdy.
\ee
The third is the power loss in the superconductor:
\be
Q_{s}=-\int \vec{J_s}\cdot\nabla V_s dxdy.
\ee

One can easily show that Eq.(9) can be obtained by variational method as a condition of minimization of the power loss $Q_1= Q_{st}+Q_{int}$:
\be
\f{\delta Q_1}{\delta V}=0,
\ee
provided that the variation $\delta V(x,y)=0$ at the boundaries and the potential $V_s$ is not varied.

If we multiply the Eq. (9) by $(V-V_s)d/\rho$, we get
\be
d/\rho \int (V-V_s)\Delta V dxdy = Q_{int}.
\ee
The integrand in the left-hand side can be rewritten as follows:
\be
(V-V_s)\Delta V = \nabla \cdot((V-V_s)\nabla V) -\nabla V\cdot\nabla (V-V_s).
\ee
At the boundaries of the conductor either $\nabla V =0$ or $V-V_s\rightarrow 0$. Therefore, taking into account Eq. (26), we obtain
\be
Q_{st}+Q_{int}= d/\rho \int \nabla V\cdot\nabla V_sdxdy\equiv
\int (\vec{J_n}\cdot\vec{E_s})dxdy.
\ee
Here $\vec{J_n}$ and $\vec{E_s}$ are the $2D$ density of current in the stabilizer and the electric field in the superconductor, respectively.

Similarly, multiplying Eq. (17) by $V-V_s$ and taking into account Eq.(28) we get
\be
Q_{s}+Q_{int}= -\int \nabla V\cdot\vec{J_s}dxdy\equiv 
\int (\vec{J_s}\cdot\vec{E_n})dxdy.
\ee
The right-hand sides of Eqs. (32) and (33) look deceptively similar to the conventional Joule heat expressions, except that the current density and electric field are from different layers. These relations may be more useful in evaluating losses in coated conductors on the basis of numerical models or magneto-optical imaging than Eqs. (26)-(28). In the areas where the superconductor and metal are equipotential, the electric field $\vec{E_s}=\vec{E_n}$ and Eqs. (32), (33) become the conventional Joule heat expressions because there is no current across the interface and no loss in the interface. 
 
\section{\label{sec:level1}Coated Conductor Stabilizer\protect}

\begin{figure}
\includegraphics{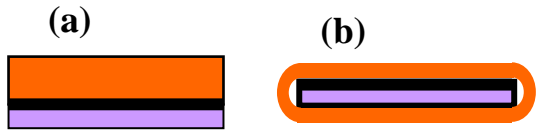}
\caption{\label{fig:} (Color online) (a) A sketch of the cross-section  of a YBCO-coated conductor with "one-sided"  stabilizer. The superconducting film (thick black line) is deposited on buffered substrate and covered with copper stabilizer. (b) "Surround stabilizer" –- a copper layer encapsulates the substrate and YBCO film. Only the top (the front) part of the stabilizer is directly attached to the superconducting film.}
\end{figure}

There are two types of stabilizers for coated conductors. The simplest, shown in Fig. 2(a), is a layer of copper of the same width as that of the YBCO film soldered or electroplated on top of the silver covered superconducting film (let us call it {\it one-sided stabilizer})\cite{Kim}. However, two of the main companies leading the development of coated superconductors (American Superconductor and SuperPower) have opted for a different type of stabilizer {\ -} {\it surround} stabilizer, shown in Fig. 2(b)\cite{Schoop,Xie}. Among the advantages of this architecture is that copper encapsulates and protects the YBCO film and the substrate. If the edges of the surround stabilizer are rounded,  it reduces the stray electric field, which is important for high voltage applications. 

Implicit in these arguments in favor of the surround stabilizer is an assumption that in terms of its ability to exchange current with the superconducting layer it is equivalent to the one-sided stabilizer with the same total thickness. This assumption needs to be critically examined.   In this section we will compare the effectiveness of these two types of stabilizers using the planar model described above. 

\subsection{\label{sec:level2} One-sided stabilizer}

A thin superconducting film consisting of a multitude of carefully aligned grains is especially vulnerable to proliferation of microcracks that limit the local critical current. Even when a newly manufactured conductor is devoid of such current limiting obstacles, they may appear as the result of stresses applied to the conductor during the coil winding and its subsequent exploitation in rotating machinery or magnets. Even a relatively short crack creates a magnetic flux jet -- a region of enhanced electric field that spans over a distance about $n—\;$times greater than the length of the crack ($n$ is the exponent in Eq. (19))\cite{Friesen}. Therefore, the hairline cracks may turn out to be one of the main types of defects that will determine the limits of reliability of the coated conductors over their life-time. 

Let us consider a specific situation when DC current flows through the coated conductor and the superconducting layer is disrupted by a hairline fracture, be it a permanent physical fracture or a linear hot spot. Then, the transport current will be diverted into the stabilizer over the fracture and then return back to the superconductor. As was discussed above, a good approximation of this situation can be a piecewise constant potential $V_s$ of the superconducting film. An implicit condition for that is that the critical current remains greater than the transport current even at elevated temperature caused by excessive dissipation.  Let the long axis of the tape be $x-$axis, and the fracture located at $x=0$, Fig. 1. Let us choose the potential of the superconducting layer $V_s=\delta V/2$ for $x>0$ and $V_s=-\delta V/2$ for $x<0$. Here $\delta V$ is the potential difference between the distant points of the conductor that is required to maintain the current across the fracture in the superconducting film. Correspondingly, the solution of Eq. (9) has the form:
\be
V=\pm\frac{ \delta V}{2}(1-e^{-\kappa |x|}).
\ee
The signs $\pm$ refer, respectively,  to $x>0$ and $x<0$. The electric field is given by
\be
E_x=-\frac{\partial V}{\partial x}=-\f{\kappa \delta V}{2} e^{-\kappa |x|}.
\ee
This is the value of the electric field at $z=0$. The electric field averaged over the thickness of the stabilizer differs by a factor $(1+d/3a)$, see Eq. (12). 

The total current in the stabilizer is given by
\be
I=\left.\f{d}{\rho}\int |E_x|dy\right |_{x=0}=
\f{\delta V Wd}{2\lambda\rho}\equiv\f{\delta V}{R},
\ee
where $W$ is the width of the conductor and
\be
R=\f{2\rho\lambda}{Wd}
\ee
is the effective resistance of the hairline fracture. Power dissipated in the stabilizer is
\be
Q_{st}=\f{Wd}{\rho}\int |E_x|^2dx=\f{1}{2}\f{\delta V^2}{R}.
\ee
Small corrections of the order of $d/a$, are neglected here. Power is also dissipated in the interface itself:
\be
Q_{int}=-\int j_z(V-V_s)dxdy=\int \f{(V-V_s)^2}{\bar{R}}dxdy=\f{1}{2}\f{\delta V^2}{R}.
\ee
Here we have used Eqs. (26) and (27).
As expected, the total power loss is simply
\be
Q_1=Q_{st}+Q_{int}=\delta VI=RI^2=\f{\delta V^2}{R}.
\ee

It is instructive to realize that half of the total energy dissipation takes place in the very thin (in comparison with the thickness of the stabilizer) interface between the superconductor and copper. In fact, in the one-dimensional case not only is the total dissipation equally divided between the stabilizer and interface, but the areal density of the power dissipation as well:
\bea
\f{d}{\rho} \left (\f{dV}{dx}\right )^2=\f{(V-V_s)^2}{\bar{R}}.\nonumber
\eea
In other words, the thermal load that the conductor experiences, is not uniformly distributed in the stabilizer. Given that the YBCO film is only about $1\mu m$ thick, practically half of all dissipated energy is deposited in the superconducting film.  This may have noticeable consequences for the dynamics of stabilization and quench development.  

The result given by Eq. (40) can be directly obtained from Eq. (32). The electric field in the superconductor with the piecewise constant potential is $|E_s| =\delta V\delta (x)$, so that from Eq. (32) follows $Q_1= \delta V I$.
For a $4\;mm$ wide coated conductor the resistance of a hairline fracture or a hot spot given by Eq.(37):
\be
R\sim 7.5\;\mu\Omega.
\ee
For the transport current $I\sim 100\;A$ the power dissipation 
\be
Q=RI^2\sim 75 \; mW.
\ee

\subsection{\label{sec:level2} Partial fracture}

For completeness, let us also consider a situation when a hairline crack in the superconducting film does not completely inhibit the flow of the transport current through the superconducting film.  In other words, let us assume that the fracture is a hairline region with the finite critical current $I_c < I$. Outside of this hairline anomaly the critical current $I_c^{\prime } > I$. This model may be considered as an approximation of a faceted grain boundary with alternating segments of different critical current density $J_c$\cite{Friesen}.

Then, over the anomaly, the excess of current $I-I_c$ will be diverted into the stabilizer.  Equations (34) and (35) are still valid, but Eq. (36) takes form
\bea
I-I_c=\left.\f{d}{\rho}\int |E_x|dy\right |_{x=0}=
\f{\delta V Wd}{2\lambda\rho}\equiv\f{\delta V}{R}.\nonumber
\eea
Power loss in the stabilizer and interface are given by Eqs. (38) and (39):
\bea
Q_{st}=Q_{int}=\f{1}{2}\f{\delta V^2}{R}=\f{1}{2}R(I-I_c)^2.\nonumber
\eea
The power dissipated in the superconducting film
\bea
Q_s = \int (\vec{J_s}\cdot\vec{E_s})dxdy = I_c\delta V =RI_c(I-I_c).\nonumber
\eea
Here we have used $|E_s|=\delta V \delta (x)$. The total power dissipation is
\bea
Q_{tot}=Q_{st}+Q_{int}+Q_{s}=\delta V I=RI(I-I_c).\nonumber
\eea
Since the thickness of the superconducting film is small in comparison to that of the stabilizer, the heat dissipated in the interface may strongly affect the temperature of the superconducting film before spreading out throughout the stabilizer. In contrast, the power dissipated in the stabilizer itself is distributed uniformly over the stabilizer volume. The amount of power dissipation that directly affects the temperature of the superconducting film is
\bea
Q_2=Q_s+Q_{int}=\f{1}{2}R(I^2-I_c^2).\nonumber
\eea
The ratio of the power dissipation in the stabilizer to that in the superconducting film is
\bea
\f{Q_{st}}{Q_2}=\f{(I-I_c)^2}{I^2-I_c^2}.\nonumber
\eea
Thus, in the case of the partial fracture the total dissipated energy is reduced in comparison with the complete fracture in proportion to $I-I_c$, but a greater part of this dissipation is strongly localized in the superconducting film. For example, if $I_c=(2/3) I$, only $20\%$ of the total dissipation is released in the stabilizer. The rest $80\%$ of heat is deposited in the superconducting film and the interface with the potentially significant consequences for the temperature distribution across the thickness of the coated conductor.

\subsection{\label{sec:level2} Surround stabilizer}
In the surround stabilizer, Fig. 2(b), only the front part of it is in electric contact with the superconducting film. The back side of the stabilizer can exchange current with the front side and with the superconductor only through the edges because the substrate is insulated from the YBCO by the buffer layer. This point is better illustrated in Fig. 3. Imagine that the back side of the stabilizer is cut along the centerline and unfolded, forming flaps.  The effectiveness of the surround stabilizer depends on how fully the current fills the back side (the flaps in Fig. 3) in the situation when the superconducting layer is disrupted as was described in the previous subsection. 
One can easily see a potential problem with the effectiveness of the flaps. Consider two points $A$ and $A^{\prime}$ separated by the distance $2\lambda$ and with potential difference between them $\delta V$. Then consider another contour $A-B-C-A^{\prime}$ that extends deep into the backside of the stabilizer. The potential difference $\delta V=-\oint \vec{E}\cdot d\vec{\ell }$ is the same along any contour. In the currently manufactured conductors the length of $ABCA^{\prime}$ contour is approximately $W+2\lambda \gg 2\lambda$ ( Eq. (23)). Therefore,  the electric field in the back side (and the density of current) must be substantially smaller than that in the front side approximately by a factor $\lambda/W$. 
\begin{figure}
\includegraphics{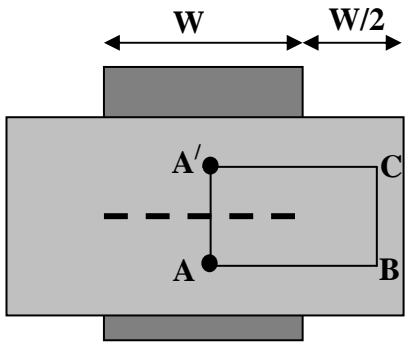}
\caption{\label{fig:} (a) A sketch of a tape-like conductor of width $W$. The back side of the stabilizer shown in Fig. 2(b) is cut along the centerline and unfolded forming two flaps $W/2$ wide each. The dashed line indicates a fracture in the superconducting film that forces transport current into the stabilizer.}
\end{figure}

Analytically, the problem of the surround stabilizer can be formulated as follows. The front side and the back side can be treated as two tape-like sheets of copper of equal thickness located within the range $-\infty <x<\infty$, $-W/2<y<W/2$. The potential in the front side $V(x,y)$ is described by Eq. (9). The potential in the back side $U(x,y)$ has to satisfy the  2D Laplace's equation:
\be
\Delta U=0.
\ee
The two solutions have to be matched at the boundaries $y=\pm W/2$:
\be
\left.\frac{\partial{V}}{\partial{x}}\right |_{y=\pm W/2}=
\left.\frac{\partial{U}}{\partial{x}}\right |_{y=\pm W/2},
\ee
and
\be
\left.\frac{\partial{V}}{\partial{y}}\right |_{y=\pm W/2}=
-\left.\frac{\partial{U}}{\partial{y}}\right |_{y=\pm W/2},
\ee
In addition, the $x=0$ (location of the fracture) must be an equipotential line with the same potential for both sides:
\be
\left. V(x,y)\right |_{x=0}=\left. U(x,y)\right |_{x=0}=const.
\ee

A complete solution of Eqs. (9) and (43)-(46) can be obtained by a straightforward, albeit cumbersome, procedure and will be presented elsewhere. However, for practical conductors this problem is characterized by a small parameter $\lambda/W\ll 1$ ($\kappa W \gg 1$). Taking advantage of this, we can find the solution perturbatively.  As an initial approximation $V^{(0)}$ we take the solution of Eq. (9) in the form (34) and (35). The solution for the  back side of the stabilizer we seek in the form
\be
U^{(1)}(x,y)=\int_{-\infty}^{\infty}U_k^{(1)}\cosh (ky)e^{ikx}\f{dk}{(2\pi )^{1/2}}.
\ee
The 	Fourier transform $U_k^{(1)}$ is determined by the boundary condition (44):
\be
\int_{-\infty}^{\infty}ikU_k^{(1)}\cosh (kW/2)e^{ikx}\f{dk}{(2\pi )^{1/2}}=
\f{\kappa \delta V}{2} e^{-\kappa |x|}.
\ee
From this follows:
\be
ikU_k^{(1)}\cosh (kW/2)= \f{ \delta V}{(2\pi)^{1/2}} \f{\kappa^2}{k^2+\kappa^2}. 
\ee
The $x-$component of the electric field $E^{(1)}_x =-\partial U^{(1)}/\partial x $ is given by:
\be
E_x^{(1)}= -\f{ \delta V}{(2\pi)^{1/2}} \int_{-\infty}^{\infty}\f{\cosh (ky)}{\cosh (kW/2)} \f{\kappa^2}{k^2+\kappa^2} e^{ikx}\f{dk}{(2\pi )^{1/2}}.
\ee
Correspondingly, the $y-$component of the electric field
\be
E_y^{(1)}= -i\f{ \delta V}{(2\pi)^{1/2}} \int_{-\infty}^{\infty}\f{\sinh (ky)}{\cosh (kW/2)} \f{\kappa^2}{k^2+\kappa^2} e^{ikx}\f{dk}{(2\pi )^{1/2}}.
\ee
\begin{figure}
\includegraphics{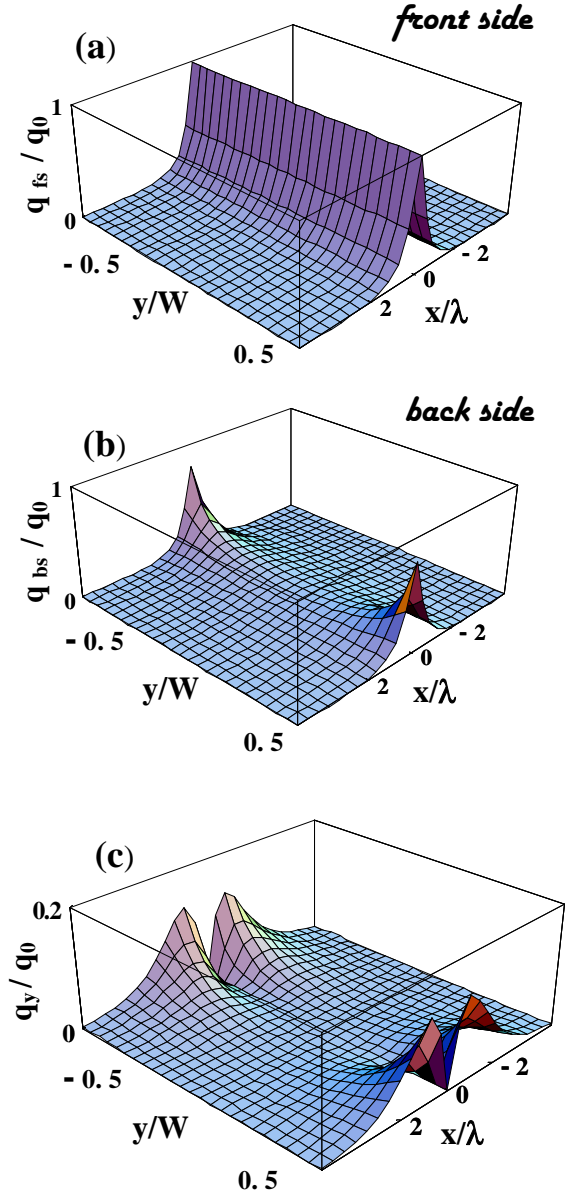}

\caption{\label{fig:} (Color online) (a) The areal density of power dissipation in the front side of the stabilizer, Eq. (53). (b) The areal density of power dissipation in the back side, Eq. (55). (c) The distribution of the transverse current, Eq. (56). The power dissipation is normalized by the value $q_0$, Eq. (54). The $x-$coordinate (along the transport current) is normalized by the screening length $\lambda$; the transverse coordinate is normalized by the width of the conductor $W$.}
\end{figure}

We can now compare the power loss in the front side of the stabilizer to the back side. The areal density of power dissipation in the front side is
\be
q_{fs}= 2 \f{d}{\rho}E_x^2. 
\ee
The factor $2$ accounts for the dissipation in the stabilizer itself, Eq. (38), and in the interface, Eq.(39). Using Eq. (35) we obtain:
\be
q_{fs}= \f{d}{\rho}\f{\delta V^2}{2\lambda^2}e^{-2\kappa|x|}.
\ee
Figure 4(a) shows the areal density of the power loss given by Eq. (53), normalized by its maximum value
\be
q_0=\f{d}{\rho}\f{\delta V^2}{2\lambda^2}.
\ee

In the back side of the stabilizer the areal density of dissipation is given by
\be
q_{bs}= \f{d}{\rho }(E_x^2+E_y^2),
\ee
where  $E_x$ and $E_y$ are given by Eqs. (50) and (51). The total amount of power dissipated in the back side and its spatial distribution, Fig. 4(b), is determined by the value of $\kappa W$. The distribution shown in Fig. 4(b) corresponds to $\kappa W=5$. For $W=4\;mm$, this is equivalent to $\lambda = 800\;\mu m$. For the value of $\lambda \approx 300\;\mu m$ measured in\cite{Polak}, $\kappa W\approx 13$, so that the amount of power dissipation in the back side is smaller than that shown in Fig. 4(b).
For comparison, in Fig. 4(c) the density of dissipation attributable to $y-$component of the current density flowing into-- and out of the back side from the front side through the edges is also shown:
\be
q_{y}= \f{d}{\rho }E_y^2.
\ee

The discontinuity in the areal density of power dissipation at the edges, $y=\pm W/2$, between the front side and the back side, Figs. 4(a,b), is not the result of approximation. The density of power dissipated in the metal itself  is continuous, by virtue of the boundary conditions (44) and (45), but additional power is dissipated in the interface in the front side. Therefore, the total dissipation in the front side is greater than that in the back side as long as there is current exchange across the interface. The difference between the amount of power dissipated in the two sides increases with increasing $\kappa W$.
 
In order to evaluate the total resistance of the hairline fracture and compare it with that for the one-sided stabilizer, Eq. (37), let us consider the total current that passes through the back side of the stabilizer: 
\be
I_{bs}=\f{d}{\rho}\int_{-W/2}^{W/2}E_xdy,
\ee
where $E_x$ is taken at $x=0$. Integrating Eq. (50) we obtain:
\be
I_{bs}=\f{2\delta V d}{\rho (2\pi)^{1/2}}\int_{-\infty}^{\infty}
\tanh (kW/2) \f{\kappa^2}{k^2+\kappa^2} \f{1}{k}\f{dk}{(2\pi)^{1/2}}
\ee
Introducing a dimensionless variable $\xi=k/\kappa $, we get:
\be
I_{bs}=\f{2\delta V d}{ \pi\rho } \int_{0}^{\infty}\tanh (\xi \kappa W/2)
\f{1}{\xi^2+1}\f{d\xi}{\xi}.
\ee
Taking into account that 
\be
\int \frac{d\xi}{\xi (1+\xi ^2)}=\ln \f {\xi}{(1+\xi^2)^{1/2}},
\ee
and integrating Eq.(59) by parts, we obtain:
\be 
I_{bs}=-\f{2\delta V d}{ \pi\rho } \f{\kappa W}{2}\int_{0}^{\infty}
\f{d\xi}{\cosh^2 (\xi\kappa W/2)}\ln \f {\xi}{(1+\xi^2)^{1/2}}.
\ee
In the limit $\kappa W \gg 1$
\bea
I_{bs}\approx -\f{2\delta V d}{ \pi\rho }\f{\kappa W}{2}\int_{0}^{\infty}
\f{ d\xi \ln \xi }{\cosh^2 (\xi\kappa W/2)}=\nonumber\\
=\f{2\delta V d}{ \pi\rho }\Big (\ln (\kappa W/2) -\Psi (1/2) -\ln\pi\Big )=\nonumber\\
=\f{2\delta V d}{ \pi\rho }\ln (1.13\kappa W).
\eea
Here $\Psi (1/2)\approx -1.96$\cite{ Grad}. The total current in the stabilizer is the sum of the current flowing in the front side, Eq. (36), and the back side, Eq. (62), is 
\be
I= \f{\delta V Wd}{2\lambda\rho}\left (1+\f{4}{\pi}\f{ \ln (1.13\kappa W)}{\kappa W}\right ).
\ee
Thus, when the screening length $\lambda \ll  W$, which is characteristic of the state of the art coated conductors, the back side of the stabilizer carries a  relatively small fraction of the total current. For example, if we take the value of $\lambda =300\;\mu m$ (Eq. (23) ) and $W=4\; mm$ ($\kappa W\approx 13$), the current in the back side is equal to a quarter of that in the front side. 

Correspondingly, the resistance of the hairline fracture stabilized by the surround stabilizer is 
\be
R_{ss}=\f{\delta V}{I}\approx \f{2\rho\lambda}{Wd}
\left (1-\f{4}{\pi}\f{\ln (1.13\kappa W)}{\kappa W}\right ).
\ee
The effectiveness of the surround stabilizer is important. In the currently manufactured coated conductors the back side takes up as much space and, therefore, decreases the engineering current density as much as the front side, but only partially contributes to stabilization of the hairline fractures. Using previous estimates for a surround stabilizer consisting of $20\mu m$ copper sheet (the total thickness of copper is $40\;\mu m $), we have to conclude that for this specific type of defects its stabilizing property (the resistance $R_{ss}$) is the same as that of  a $25\mu m$ one sided stabilizer (Eq. (37)). 

This conclusion does not apply to the situation when the length of the normal zone is greater or of the order of the width of the conductor. Then, the current flowing in the stabilizer fills more or less uniformly both sides of the stabilizer. However, this corresponds to quench when stability has been already breached. It should me emphasized that in most experiments in which the stability of coated conductors has been tested so far, the dominant scenario is the creation of the initial normal zone of a fairly large area\cite{Wang}. In these situations there is little difference between the stabilizing properties of the one-sided and surround stabilizers. The effectiveness of the stabilizer with respect to the initial hairline fracture has not been addressed yet.

\section{\label{sec:level1} Summary and suggestion }
The planar model that describes the current sharing between the superconducting film and adjacent metal, including the expressions for energy dissipation (33) and (34), may become a useful addition to the powerful analytical\cite{Friesen} and numerical methods that have been developed to study the current distribution in superconducting films, allowing to expand them to practical coated conductors. One noteworthy conclusion of our analysis is that when the dc current is forced from the superconductor into stabilizer, as much as half of the total energy dissipation takes place in the very thin interface on the surface of the superconducting film, Eq. (39).  

The surround stabilizer is not as effective as the one sided stabilizer in the case of hairline fracture if the screening length is small in comparison to the width of the conductor. Therefore, in order to stabilize coated conductors against all potential perturbations, including the hairline cracks, the total thickness of copper in the surround stabilizer would have to be greater than that in the one-sided stabilizer. Since high engineering current density is a major potential competitive advantage of the coated conductors over conventional copper wires, this shortcoming needs to be corrected. An obvious solution is to use asymmetric surround stabilizer, where the front side has greater thickness than the back side. This retains the advantages of the surround stabilizer configuration such as encapsulation of the YBCO film by copper and the possibility to round the edges. At the same time the total thickness of copper can be made smaller because the main contribution to stabilization is provided by the front part of the stabilizer. An additional bonus of the asymmetric architecture of the surround stabilizer is that it places the YBCO film closer to the neutral axis.

\begin{acknowledgments}
We thank T. Haugan and C. Varanasi for useful discussions.
\end{acknowledgments}                                                                                                                                                      

%\bibliography{apssamp}

\begin{references}
\bibitem{Larbalestier} D. Larbalestier, A. Gurevich, D. M. Feldman, and A. Polyanskii, Nature 414, 368 (2001)
\bibitem{Schoop} Schoop, U.; Rupich, M.W.; Thieme, C.; Verebelyi, D.T.; Zhang, W.; Li, X.; Kodenkandath, T.; Nguyen, N.; Siegal, E.; Civale, L.; Holesinger, T.; Maiorov, B.; Goyal, A.; Paranthaman, M. IEEE Trans. Appl. Supercond. 15, 2611-2616 (2005).
\bibitem{Xie} Y.-Y. Xie, A. Knoll, Y. Chen, Y. Li, X. Xiong, Y. Qiao, P. Hou, J. Reeves, T. Salagaj, K. Lenseth, L. Civale, B. Maiorov, Y. Iwasa, V. Solovyov, M. Suenaga, N. Cheggour, C. Clickner, J.W. Ekin, C. Weber and V. Selvamanickam,  Physica C 426-431, 849-857 (2005).
\bibitem{Iijima}Y. Iijima, K. Kakimoto, Y. Sutoh, S. Ajimura, and T. Saitoh, IEEE Trans. Appl. Supercond. 15, 2590 (2005).
\bibitem{Usoskin1} A. Usoskin, A. Rutt, J. Knoke, H. Krauth, and Th. Arndt, IEEE Trans. Appl. Supercond. 15, 2604 (2005).
\bibitem{Angurel} L. A. Angurel, M. Bona, J M Andr´es, D Mu˜noz-Rojas. and
N Casa˜n-Pastor,  Supercond. Sci. Technol. 18, 135 (2005) 
\bibitem{Fang}Y. Fang, S. Danyluk, Y. S. Cha, and M. T. Lanagan, J. Appl. Phys. 79,  947 (1996)
\bibitem{Polak2} M. Polak, W.J. Parrell, X. Y. Cai, A. Polyanskii, E. E.Hellstrom, D. C. Larbalestier, and M.Majoros,  Supercond. Sci. Technol. 10 (1997) 769-777 and references therein.
\bibitem{Cesnak} L. Cesnak, P. Kovac, and F. Gomory, Supercond. Sci. Technol. 13 (2000) 1450
\bibitem{Usoskin} A. Usoskin, A. Issaev, H.C. Freyhardt, M. Leghissa, M.P. Oomen, H.-W. Neumueller, Physica C 372-376 (2002) 857-862
\bibitem{Fu} Y. Fu, O. Tsukamoto, and M. Furuse, IEEE Trans. Appl. Supercond,  13,  1780 (2003)
\bibitem{Stenvall} A. Stenvall, A. Korpela, J. Lehtonen, and R. Mikkonen, Supercond. Sci. Technol. 20 (2007) 92 and references therein.
\bibitem{Takacs} S. Tak\`acs, Supercond. Sci. Technol. 20, 180 (2007) 
\bibitem{Friesen} M. Friesen and A. Gurevich, Phys. Rev. B 63, 064521 (2001).
\bibitem{Polak} M. Polak, P. N. Barnes,  and G. A. Levin,  Supercond. Sci. Technol. 19 (2006) 817
\bibitem{Kim} K. Kim, M. Paranthaman, D. P. Norton, T. Aytug, C. Cantoni, A. A. Gapud, A. Goyal, and D. K. Christen, Supercond. Sci. Technol. 19 (2006) R23
\bibitem{Grad} I. S. Gradshteyn and I. M. Ryzhik, Table of Integrals, Series and Products, Academic Press, San Diego, 1980.
\bibitem{Wang} X. Wang, U. P. Trociewitz, and J. Schwartz, J. Appl. Phys. 101, 053904 (2007) and references therein

\end{references}

.\\

%{\bf Captions to Figures:}\\
%{\bf Fig.1 (a):} 
\newpage
\end{document}